\documentclass[aps,prx,twocolumn,floatfix,superscriptaddress]{revtex4-2}

\usepackage{amssymb,amsmath,amstext}                
\usepackage{graphicx}
\usepackage{tabularx}
\usepackage{booktabs}
\usepackage{lipsum}
\usepackage{epstopdf}                                               
\usepackage{color}                                                     
\usepackage{bm}                                                        
\usepackage{appendix}
\usepackage[utf8]{inputenc}
\usepackage{lipsum}
\usepackage{tikz,xcolor}
\usepackage{ulem}
\usepackage{latexsym}
\usepackage[colorlinks=true,citecolor=blue,linkcolor=magenta]{hyperref}
\usepackage{cleveref}
\usepackage{soul}
\usepackage{amsmath}
\usepackage{float}
\usepackage{siunitx}
\usepackage{comment}

\def\be{\begin{equation}}
\def\ee{\end{equation}}
\def\bea{\begin{eqnarray}}
\def\eea{\end{eqnarray}} 
\def\ra{\rangle}

\def\bi{\begin{itemize}}
\def\ei{\end{itemize}}

\definecolor{dgreen} {RGB}{78,138,21}

\definecolor{giergiel} {RGB}{0,128,88}

\definecolor{purple} {RGB}{128,0,160}

\usepackage{orcidlink}

\begin{document} 
\title{Large Anomalous Shifts of Potassium-39 Feshbach Resonances}

\author{Ali Zaheer\,\orcidlink{X}}
\affiliation{Optical Sciences Centre, Swinburne University of Technology, Melbourne, Victoria 3122, Australia} 

\author{Mohammed Bouras\,\orcidlink{X}}
\affiliation{Optical Sciences Centre, Swinburne University of Technology, Melbourne, Victoria 3122, Australia}

\author{Krzysztof Giergiel\,\orcidlink{X}}
\affiliation{Optical Sciences Centre, Swinburne University of Technology, Melbourne, Victoria 3122, Australia}
\affiliation{CSIRO, Technology, Research Way, Clayton, Victoria 3168, Australia}
\affiliation{Instytut Fizyki Teoretycznej, Uniwersytet Jagiello\'{n}ski, ulica Profesora Stanislawa Lojasiewiewicza 11, PL-30-348 Krak\'{o}w, Poland}
\author{Rudolf Grimm\,\orcidlink{X}}
\affiliation{Institut f\"ur Quantenoptik und Quanteninformation (IQOQI), \"Osterreichische Academie der Wissenschaften, 6020 Innsbruck, Austria}
\affiliation{Institut f\"ur Experimentalphysik, Universit\"at Innsbruck, 6020 Innsbruck, Austria}

\author{Andrei Sidorov\,
\orcidlink{X}},
\email{asidorov@swin.edu.au}
\affiliation{Optical Sciences Centre, Swinburne University of Technology, Melbourne, Victoria 3122, Australia}

\author{Peter Hannaford\,
\orcidlink{X}},
\email{phannaford@swin.edu.au}

\affiliation{Optical Sciences Centre, Swinburne University of Technology, Melbourne, Victoria 3122, Australia}

\begin{abstract}
We report the observation of large anomalous shifts, up to +7.5 G, of the positions of the 33.6 G and 39.9 G Feshbach resonances in potassium-39 atoms confined in a 1063.9 nm optical dipole trap (ODT) at temperatures up to around 35 $\mu$K and trap depths up to about 136 $\mu$K. When the atom cloud is cooled to lower temperatures, by reducing the trap depth of the ODT, the shifts decrease proportionally with trap depth and approach zero at zero depth. We show that the large observed shifts originate from a large differential ac Stark shift between the incoming pair of potassium-39 atoms and the weakly bound Feshbach molecule, which in turn originates from an unexpectedly large dynamic polarizability of the Feshbach molecule. The polarizabilities of the Feshbach molecules extracted from the measured shifts of the 33.6 G and 39.9 G resonances are about four to seven times the sum of the polarizabilities of the two incoming potassium-39 atoms, that is, about four to seven times larger than the usual polarizability of weakly bound Feshbach molecules. The large polarizabilities of the Feshbach molecules are attributed to a near-coincidence between the frequency of the 1063.9 nm ODT laser and the frequency of a molecular transition from the last vibrational level of the lowest triplet $a^3\Sigma_u^+$ potential to a vibrational level of the excited $b^3\Sigma_g^+$ potential. Other potassium-39 Feshbach resonances we have studied exhibit a zero or very small shift, corresponding to molecular polarizabilities close to the sum of the polarizabilities of the two incoming potassium-39 atoms.
\end{abstract}

\date{\today}
\maketitle

\section{Introduction}

Since the pioneering work on ultracold potassium atoms in the late 1990s \cite{Wang96,Cataliotti98,DeMarco99} and the first realization of  a potassium-39 Bose-Einstein condensate (BEC) in 2007 \cite{Roati07}, potassium-39 has been widely used by many groups to create ultracold samples of bosonic atoms with precisely tuneable inter-atomic interactions \cite{Roati07, Derrico07, Campbell10, Landini12, Marangoni12, Salomon14a, Roy13, Nath13, Salomon14b, Jorgensen16, Culver16, Tanzi18, Schulze18, Chapurin19, Wang20, Herbst22, Yu24, Zhang25}. Potassium-39 has a number of broad $s$-wave Feshbach resonances that enable the scattering length to be precisely tuned over a wide range of positive and negative values, including zero. Examples of applications of ultracold $^{39}\textup{K}$ atoms with tuneable interactions include studies of low-dimensional Bose gases \cite{Fletcher15}, unitary Bose gases \cite{Fletcher13}, Efimov trimer states \cite{Roy13, Chapurin19}, attractive and repulsive polarons \cite{Jorgensen16}, high precision atom interferometry and gravimetry \cite{Petrucciani25}, and quantum information processing \cite{Wang20}.

The first experimental and theoretical studies of $^{39}\textup{K}$ Feshbach resonances were reported by the Florence\ /Rennes groups \cite{Derrico07}, which included eight $s$-wave resonances involving single-spin pairs of the atomic states $|F=1, m_F=+1\ra$, $|F=1, m_F=0\ra$ and $|F=1, m_F-1\ra$. More recently, the experimental positions of the $^{39}\textup{K}$ Feshbach resonances have been updated \cite{Roy13} and are now in excellent agreement with the theoretical values based on coupled-channel calculations \cite{Derrico07, Lysebo10, Tiemann20}, see Table I.   

Most of the studies using $^{39}\textup{K}$ Feshbach resonances have involved the high magnetic field $s$-wave resonance at 402.6 G \cite{Chin10}. In recent years, the $s$-wave resonance at 33.6 G has become more widely used to create $^{39}\textup{K}$ BECs \cite{Herbst22, Herbst24, Hammond23} due to the  readily accessible magnetic field, which is convenient for compact cold-atom experiments, such as experiments conducted in space \cite{Herbst24}. The 33.6 G resonance is a suitably broad (width $\Delta$ = 55 G [5]) intermediate-strength Feshbach resonance, which allows precise tuning of the scattering length, and originates from a low magnetic field seeking atomic state $|F=1, m_F=-1\ra$, thus allowing the atoms to be magnetically trapped or confined in a magnetic field gradient, if required. On the other hand, the 33.6 G resonance is not suitable for creating scattering lengths close to zero owing to a small overlap with the oppositely-poled $|F=1, m_F=-1\ra$ resonance at 162.3 G \cite{Derrico07}, see Fig. 1. 

\begin{table*}
\begin{center}
\caption{Potassium-39 Feshbach resonances studied in this work, with reported parameters. $a_{bg}$ is the background scattering length. $\delta \mu = \mu_{oc} - \mu_{cc}$ is the differential magnetic moment, where $oc$ is the open (atomic) channel, $cc$ is the  closed (molecular) channel, and ${\mu_B}$ is the Bohr magneton. $s_{res}=(a_{bg}\Delta/\bar{a})(\delta\mu/\bar{E})$ is the resonance strength \cite{Chin10}, where $\bar{a}= [4\pi/\Gamma(\frac{1}{4})^2]r_{vdW}=0.955978...r_{vdW}$, $\bar{E}=\hbar^2/(2\mu_{12}\bar{a}^2)=1.09422...E_{vdW}$, $r_{vdW}=\frac{1}{2} (\frac{2\mu C_6}{\hbar^2})^{1/4}$, $E_{vdW}=\frac{\hbar^2}{2\mu_{12}}\frac{1}{r^2_{vdw}}$, and $\mu_{12}$ is the reduced mass. Atom-pair spin channels: $\{aa\}$=$|1, +1\ra$; $\{bb\}$=$|1, 0\ra$; $\{cc\}$=$|1, -1\ra$; $\{ac\}$=$|1, +1\ra+|1, -1\ra$; $\{ab\}$=$|1, +1\ra+|1, 0\ra$; $\{bc\}$=$|1, 0\ra+|1, -1\ra$. [tw] refers to this work.
}
\vspace{2mm}
\begin{tabularx}{\textwidth}{|c|c|c|c|c|c|c|c|c|}
\hline
\hline
Label & Position $B_0$(G) & $B_0$(G) & \@ \@ Width $\Delta$(G) \@ \@  &  $a_{bg}/a_0$  & \@  Atom-pair \@  & Molecular state & Diff. magnetic & \@ \@ \@ Strength \@ \@ \@ \\
& Experiment & Theory & Theory & \@ \@ \@ Theory\@ \@ \@  & spin channel & ($S,I,f,M$) & moment $\delta \mu/\mu_B$ & $s_{res}$ [tw] \\
\hline
R0A & 25.85(10) \cite{Derrico07} & 25.9 \cite{Derrico07} & $-0.47$\cite{Derrico07} & $-33$\cite{Derrico07} &\{$aa$\} & (1, 3, 3, +2) & $+1.5$ \cite{Derrico07} & 0.022 \\
 &  & 25.90 \cite{Lysebo10} & $-0.4315$\cite{Lysebo10} & $-35.73$ \cite{Lysebo10} &  & & $+1.6$ [tw] &\\
 &  &  25.874 \cite{Tiemann20} & & & & &  &\\
\hline
R0B &  & 25.8 \cite{Lysebo10} & $-1.252$\cite{Lysebo10} & $-38.02$\cite{Lysebo10} & $\{ab\}$ & (1, 1, 1, +1) &  +1.6 [tw] & 0.06 \\& & & & & & & &\\
\hline
R1 & 33.64(15) \cite{Roy13} & 33.6 \cite{Derrico07} & +55 \cite{Derrico07} & $-19$ \cite{Derrico07} & \{$cc$\} & (1, 1, 2, $-2)$ & $-1.9$ \cite{Derrico07}  & 1.9 \\
 & 33.5820(14) \cite{Chapurin19} & 33.61 \cite{Lysebo10} & +79.82\cite{Lysebo10} & $-13.46$\cite{Lysebo10} & &  & $-2.5$ [tw] &  \\
 &  &  33.582 \cite{Tiemann20} & +54.772 \cite{Chapurin19} & $-19.599$\cite{Chapurin19} & & & &\\
\hline
R3 &   & 39.86 \cite{Lysebo10} & $-2.196$\cite{Lysebo10} & $-38.46$\cite{Lysebo10} & \{$ab$\} & (1, 3, 3, +1)  & +1.2 [tw] & 0.09 \\
& & & & & & & &\\
\hline
R4 & & 52.6 [tw] & & & $p$-wave \{$bc$\}\  & & &  \\ 
\hline
R5 & 58.92(3) \cite{Roy13} &  58.8 \cite{Derrico07} & $-9.6$\cite{Derrico07} & $-18$\cite{Derrico07} & \{$bb$\} & (1, 3, 3, 0) & $+0.83$\cite{Derrico07} & 0.14 \\
  &  & 58.86 \cite{Lysebo10} & $-10.45$\cite{Lysebo10} & $-15.92$\cite{Lysebo10} &  & & $+0.8$ [tw] & \\
  &  &  59.62 \cite{Tiemann20} & & & & & &\\
\hline
R6 & 65.67(5) \cite{Roy13} & 65.6 \cite{Derrico07} & $-7.9$\cite{Derrico07} & $-18$\cite{Derrico07} & 
\{$bb$\} & (1, 1, 1, 0) & +0.78 \cite{Derrico07} & 0.11\\
 &  & 66.49 \cite{Lysebo10} & $-5.403$ \cite{Lysebo10} & $-21.08$ \cite{Lysebo10} & & & $+1.0$ [tw] &  \\
 &  &  65.57 \cite{Tiemann20}& & & & & &\\
\hline
\end{tabularx}
\end{center}
\end{table*}

In recent studies of the 33.6 G Feshbach resonance, we have found that the position of the Feshbach atom-loss resonance can shift to higher magnetic fields by as much as 7.5 G for $^{39}\textup{K}$ atoms confined in a 1063.9 nm optical dipole trap (ODT) at temperatures up to around 35 $\mu$K and trap depths up to  about 136 $\mu$K. When the trapped atoms are evaporatively cooled to lower temperatures by reducing the trap depth of the ODT, the shift of the Feshbach resonance decreases proportionally with trap depth and approaches zero at zero depth. The observed shift is more than an order-of-magnitude larger than, and of opposite sign to, the estimated temperature-induced shift of $-$0.009 G/$\mu$K for the 33.6 G resonance, based on a differential magnetic moment of $-2.5 \mu_B$. The large observed shift is attributed to a large differential ac Stark shift similar to, but much larger than, the differential ac Stark shifts observed for various single atomic species \cite{Bauer09,Vexiau11} and dual species \cite {Kohstall12, Cetina15, Peng18, Gregory17, Lous18, Khlebnikov21} in optical dipole traps and for dual species \cite{Spence23, Wei26} tightly confined in optical tweezers. We have found a similarly large shift for the $^{39}\textup{K}$ 39.9 G mixed-spin $|1, +1\ra+|1, 0\ra$ Feshbach resonance. Such large and variable shifts of Feshbach resonances can compromise precision measurements in optical dipole traps \cite{Kohstall12} and can complicate identification and implementation of the 33.6 G Feshbach resonance for creating a $^{39}\textup{K}$ BEC in an ODT. Such shifts may also have useful application, for example, in ultrafast optical tuning of Feshbach resonances, on time scales down to hundreds of nanoseconds \cite{Cetina15}. 

In this paper, we report a detailed study of the  large shifts of the $^{39}\textup{K}$ 33.6 G and 39.9 G Feshbach resonances in an atom cloud evaporatively cooled in a 1063.9 nm optical dipole trap, along with the shifts of six other $^{39}\textup{K}$ Feshbach resonances (Table I). Of these, only the 33.6 G and 39.9 G resonances exhibit large shifts.  

The paper is structured as follows. In Sec. II, we present calculations of the $^{39}\textup{K}$ Feshbach resonances up to 200 G and the relevant Zeeman molecular energy levels for eight Feshbach resonances in  the $^{39}\textup{K}_2$ molecule. In Sec. III we describe our experimental setup for measuring the positions of the resonances, and in Sec. IV we present the main experimental results, including the measured shifts of the resonances. In Sec. V, we discuss the origin of the large polarizabilities of the molecular states associated with the $^{39}\textup{K}$ 33.6 G and 39.9 G Feshbach resonances, and in Sec. VI we summarize our results and conclusions.

\section{$^{39}\textup{K}$ Feshbach Resonances}
\label{tpcb}
  
 The $^{39}\textup{K}$ Feshbach resonances in this study involve coupling between atom-pair hyperfine states and molecular hyperfine states of the same magnetic quantum number $M$ within the last vibrational level in the triplet $a^3\Sigma_u^+$ potential. The relevant triplet $f$=1 molecular hyperfine states are the $(S, I, f)$ = $(1, 3, 3)$, $(1, 1, 1)$, and $(1, 1, 2)$ states, where $S$, $I$ and $f$ are the molecular electron-spin quantum number, the nuclear-spin quantum number, and the total molecular hyperfine angular momentum quantum number, respectively \cite{Derrico07, Lysebo10, Tiemann20}. 
 
 A two-body, coupled-channel (CC) scattering model \cite{Derrico07} involving the use of optimized adiabatic Born-Oppenheimer singlet $X^1\Sigma_g^+$ and triplet $a^3\Sigma_u^+$ interaction potentials determined from spectroscopic data \cite{Amiot95, Zhao96} is employed to calculate $\textup{K}_2$ interatomic potentials for the molecular states associated with the Feshbach resonances \cite{Simoni26}. The interaction potentials at large internuclear separations are parameterized in terms of $s$-wave singlet and triplet scattering lengths, $a_S$ = 138.90(15)$a_0$ and $a_T$ = $-33.3(3)a_0$, and the long-range van der Waals coefficient, $C_6$ = 3921(8) a.u., where $a_0$ is the Bohr radius. 

Figure 1 shows the calculated $^{39}\textup{K}$ Feshbach resonances for $F$ = 1 $s$-wave and $d$-wave resonances up to 200 G, where the uncertainties are of the order of 0.1 G \cite{Simoni26}. $S$-wave resonances occur at 25.9 G, 33.6 G, 39.9 G, 58.8 G, 65.6 G, 69.2 G, 77.8 G, 113.9 G and 162.3 G. The calculations also predict some very narrow $d$-wave resonances, including one at 60.5 G (see also \cite{Lysebo10}). The  calculated center positions $B_0$ of the resonances are in excellent agreement with previously reported theoretical positions \cite{Derrico07, Lysebo10, Tiemann20} and the updated experimental positions of the resonances \cite{Roy13, Chapurin19} (Table I).

\begin{figure*}[t]
\centering
\hspace*{1.3cm}
\includegraphics[width=0.87\textwidth]{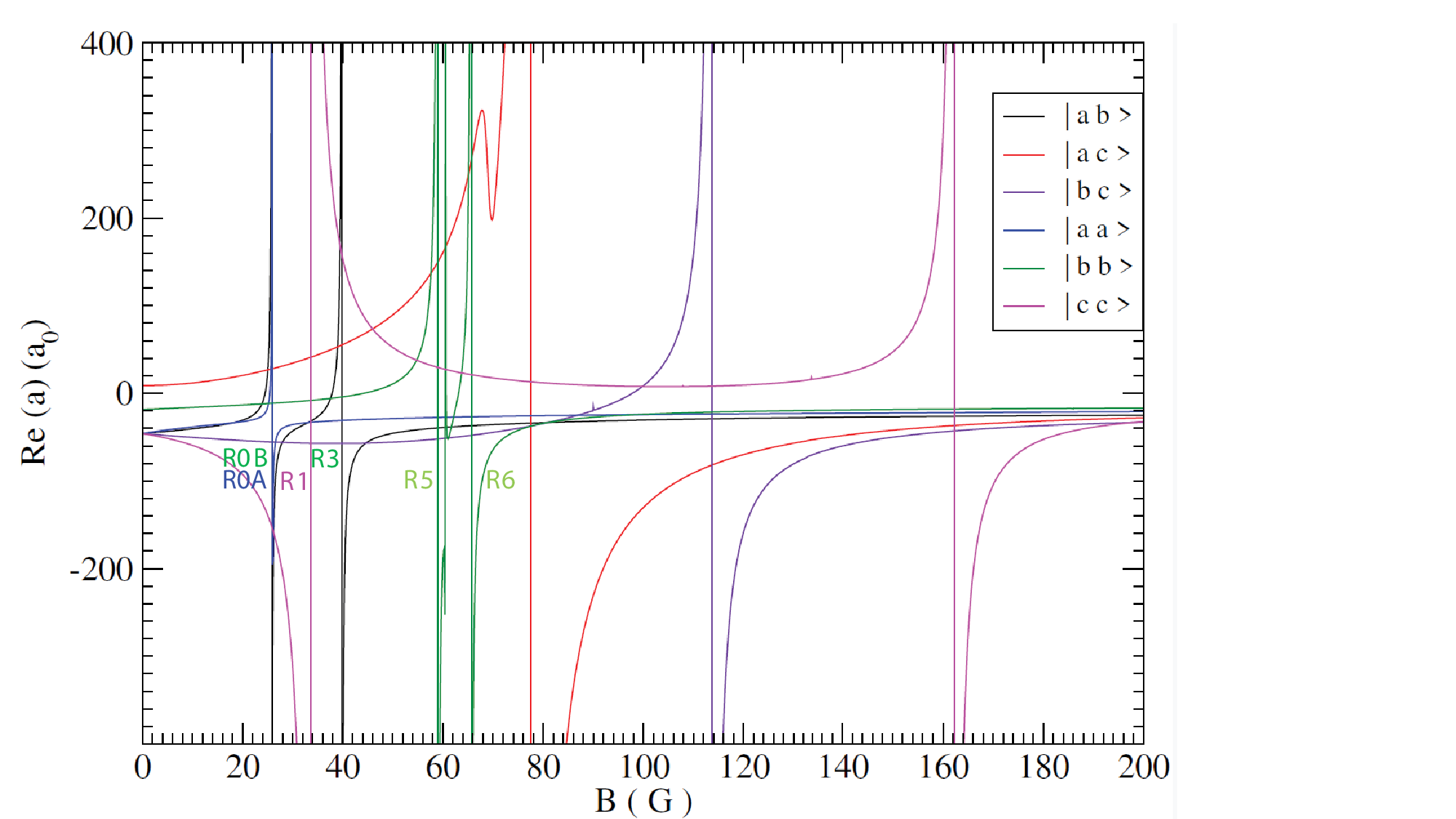}
\caption{Calculated $^{39}\textup{K}$ $F$=1 $s$-wave and $d$-wave Feshbach resonances up to 200 G \cite{Simoni26} . The spin channels $\{aa\}$, etc, are defined in the caption to Table I. The resonances studied in this work are labeled R0A, R0B, R1, R3, R5, and R6 (Table I).}
\label{figure*}
\end{figure*}

\begin{figure*}[t]
\centering
\hspace*{0.60cm}
\includegraphics[width=0.77\textwidth]{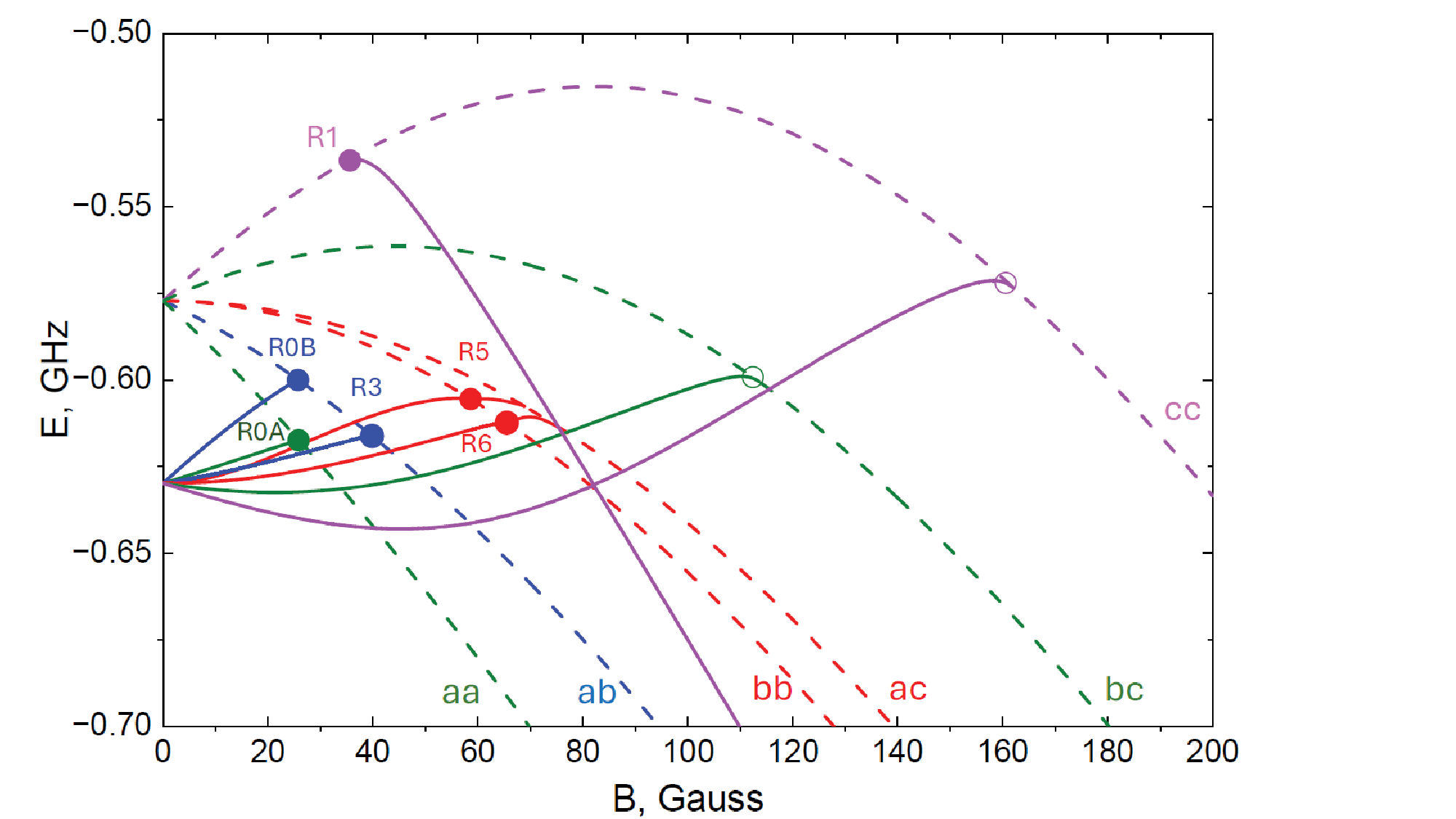}
\caption{Calculated partial molecular and atom-pair Zeeman energy-level diagram leading to $F$=1 $s$-wave Feshbach resonances at 25.9 G (R0A and R0B), 33.6 G (R1), 39.9 G (R3), 58.8 G (R5), 65.6 G (R6), 77.8 G, 113.9 G and 162.3 G \cite{Simoni26}, where the energy of atom-pair states (dashed lines) equals the energy of bound molecular states (solid lines) and the magnetic quantum numbers $m_{F1} + m_{F2} = M$. The molecular Zeeman states are $(S, I, f, M)=(1, 3, 3, +2)$ [dark green], (1, 1, 1, +1) [blue], (1, 3, 3, +1) [blue], (1, 1, 1, 0) [red], (1, 3, 3, 0) [red], (1, 3, 3, -1) [light green], (1, 1, 2, -2) [purple, 33.6 G resonance], (1, 1, 2, -2) [purple, 162.3 G resonance]. The (1, 3, 3, $M$) and (1, 1, 1, $M$) states are degenerate in zero magnetic field, as described in text.}     
\label{Figure*}
\end{figure*}

Figure 2 shows the calculated Zeeman energy-level curves for the relevant $F$=1 atom-pair magnetic states ($m_1, m_2$) and the molecular magnetic states $(1, 3, 3, M)$, $(1, 1, 1, M)$, and $(1, 1, 2, -2)$ within the triplet $a^3\Sigma_u^+$ manifold \cite{Simoni26}. We note that the pair of molecular hyperfine states $(1, 3, 3)$ and $(1, 1, 1)$ represent a special case in which $f=I$ and the same $S$. In this case, the zero-field energies of the molecular states, which are proportional to $f(f+1)-S(S+1)-I(I+1)$ for electron-spin dominated $^3\Sigma$ states, reduce to $-S(S+1)$, which for the $(1, 3, 3)$ and $(1, 1, 1)$ states, leads to a remarkable degeneracy at zero magnetic field. Also, for states with $f=I$ and the same $S=1$, the Land\'e factor $g_f$ is proportional to $1/[f(f+1)]$; so that the Zeeman shift for the (1, 1, 1, +1) state at weak magnetic fields can theoretically be six-times larger than for the (1, 3, 3, +1) state. This factor, which reduces to about three as a result of non-adiabatic singlet-triplet mixing, accounts for the large Zeeman shift for the upper of the two $M = +1$ states, leading to the R0B resonance in Fig. 2. This large Zeeman shift allows the upper $M = +1$ state to be identified as the (1, 1, 1, +1) state and the lower $M = +1$ state as the (1, 3, 3, +1) state.

$S$-wave Feshbach resonances occur at magnetic-field values where the energy of the incoming pair of $^{39}\textup{K}$ atoms equals the energy of a weakly bound molecular state having the same magnetic quantum number, i.e., $m_1 + m_2 = M$ (Fig. 2, circles).  According to Fig. 2, the resonances labeled R0A, R3 and R5 originate from the $(S, I, f)$=(1, 3, 3) molecular state, R0B and R6 from the (1, 1, 1) state, and  R1 from the (1, 1, 2) state. As a result, R0, R3, R5, and R6 all have similar energies at the Feshbach resonance, while R1 is separated from the other resonances by the $f$=2 to $f$=3 molecular hyperfine plus Zeeman splitting of about 80 MHz. In the case of R1, the resonance occurs at small positive energy where the molecular state is a quasi-bound state (Fig. 2).

A summary of the various $^{39}\textup{K}$ Feshbach resonances studied in this work is presented in Table I along with reported parameters. We note that the theoretical values for the width $\Delta$ and the background scattering length $a_{bg}$ for the R1 resonance in \cite{Lysebo10} differ significantly from those in \cite{Derrico07} and \cite{Chapurin19}. Close to the Feshbach center, the (diverging) scattering length can be written as $a=a_{bg} \Delta/ \delta B$, where the product $a_{bg} \Delta$ has essentially the same value in all three papers. The differences for the individual values of $\Delta$ and $a_{bg}/a_0$ in \cite{Lysebo10} most likely arise from the overlap between the R1 $\{cc\}$ resonance and the oppositely poled 162.3 G $\{cc\}$ resonance (Fig. 1). Similar differences for $\Delta$ and $a_{bg}$ in \cite{Lysebo10} occur for the 162.3 G resonance. 

The strength of the Feshbach resonance $s_{res}$, which is proportional to $a_{bg} \Delta /\delta \mu$ (and defined in Table I), characterizes the coupling strength between the open (free-atom) channel and the closed (bound-dimer state) channel. All resonances, apart from R1, have a coupling strength $s_{res} \ll 1$, indicating that these are closed-channel dominated resonances, for which the wavefunction of the atom pair predominantly has the character of a bound dimer. The broad R1 resonance has intermediate strength ($s_{res}$ = 1.9) between a closed-channel dominated and an open-channel dominated resonance, featuring substantial contributions by the bound dimer state and the scattering state. 

The slopes of the Zeeman molecular and atom-pair energy curves close to the Feshbach resonance magnetic fields in Fig. 2 allow a determination of the differential magnetic moment $\delta \mu = \mu_{oc} - \mu_{cc}$ between the atomic state (open channel, $oc$) and the molecular state (closed channel, $cc$) for each resonance (Table I). This method works well for the narrow, weak resonances ($s_{res} \ll 1$) in which the Zeeman molecular energy curves cross the Zeeman atom-pair energy curves in the linear regions of both curves, as for R0A, R0B, R3 and R6. However, for the broad, intermediate-strength resonance R1, the slope of the apparent linear region can lead to systematic error in the measured molecular magnetic moment at the Feshbach resonance center, which could account for the difference between the value reported in \cite{Derrico07} and the value obtained in this work (Table I). Also, in the case of R5, the crossing of the Zeeman molecular energy curve with the atom-pair curve in the strongly nonlinear region can lead to systematic error. 

We note that the differential magnetic moment for the R1 resonance is large and negative, while it is positive for the other Feshbach resonances studied. The negative $\delta \mu$ for R1 arises because this resonance is associated with the $f$=2, $I$=1 molecular hyperfine state, which is located 142 MHz above the $F$=1 atom-pair hyperfine state at zero magnetic field, whereas the other resonances are associated with the (degenerate) $f$=3, $I$=3 or $f$=1, $I$=1 molecular hyperfine states, which are located 53 MHz below the $F$=1 atom-pair hyperfine state at zero magnetic field (Fig. 2). 

\section{Experimental Arrangement}
\label{qcsec}

We use an all-optical multi-stage set-up to cool the $^{39}\textup{K}$ atom cloud, similar to that reported by Salomon et al. \cite{Salomon14b} and Herbst et al. \cite{Herbst22}.  

Briefly, about $10^9$ cold $^{39}\textup{K}$ atoms from a beam emanating from a 2D magneto-optical trap (MOT) are loaded into a 3D MOT, operating on the potassium 767 nm (D2) $4S_{1/2}$, $F$=2 $\rightarrow$ $4P_{3/2}$, $F'$=3 cooling transition and the $4S_{1/2}$, $F$=1 $\rightarrow$ $4P_{3/2}$, $F'$=2 repumping transition, where the atom cloud is cooled to around 2 mK, which is limited by the narrow hyperfine splitting in the $4P_{3/2}$ excited state. The atom cloud is then compressed and cooled in an intermediate hybrid MOT stage, consisting of blue-detuned 770 nm (D1) cooling light and red-detuned 767 nm (D2) repumping light, to a temperature of around 170 $\mu$K. The D2 repumping beam and D1 cooling beam are then ramped down to zero intensity and about 30\% of the saturation intensity, respectively, to pump atoms into the $F$=1 ground state. In the next stage, the atom cloud is cooled in a blue-detuned D1 gray molasses [9], in which the frequency difference between the cooling and repumping D1 beams from the same laser is set to the $4S_{1/2}$ ground-state hyperfine splitting of 461.72 MHz, via a pair of acousto-optic modulators, to reach temperatures down to 6 - 10 $\mu$K.

Next, about $3 \times 10^5$ $^{39}\textup{K}$ atoms from the gray molasses are loaded into a crossed optical dipole trap (ODT) formed from a free-running single-frequency 1063.9 nm fiber laser (Azur Light Systems ALS-IR-1064-50-I-CC-SF, linewidth $<50$ kHz) with a measured frequency of 281.7665(2) THz. The frequency of the fiber laser could be shifted by +13.1 GHz by allowing a stable temperature-dependent mode hop. The incoming primary beam (14 W maximum power) is focused into a 200 $\mu$m waist where it crosses the secondary beam (11 W maximum power) with 95 $\mu$m waist at an angle of $70^\textup{o}$ in a bow-tie configuration. The primary beam is directed at Brewster’s angle to a UHV glass cell with horizontal polarization, while the secondary beam has vertical polarization parallel to the vertical magnetic field. The trap depth of the secondary beam is about four times that of the primary beam, so the ODT beam has predominantly linear polarization along the vertical direction. The trap frequencies along the three axes were measured to be (121, 511, 523) Hz at the maximum power (where 523 Hz is the vertical confinement), using induced dipole oscillations. A maximum trap depth of 136 $\mu$K was determined for the maximum powers in the two beams.

\begin{figure}[t]
\centering
\includegraphics[width=0.68\textwidth]{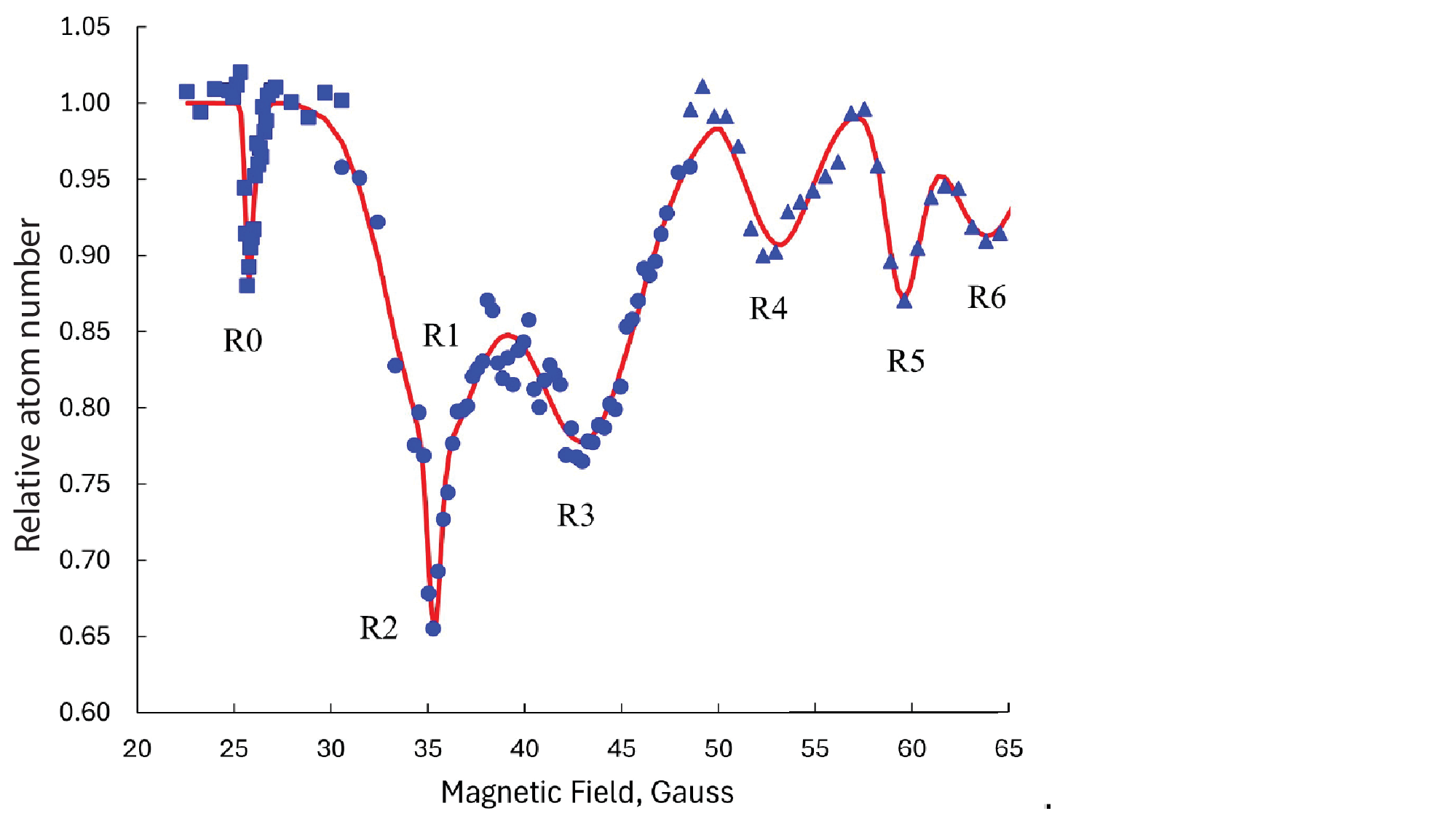}
\caption{Scan of the Feshbach atom-loss resonances up to 65 G for a $^{39}\textup{K}$ unpolarized-spin sample at a temperature of 16 $\mu$K and ODT trap depth of 46 $\mu$K. The blue squares, circles and triangles represent data taken on different days. The solid line represents a multi-Gaussian fit. The labels are defined in Table I.} 
\label{Figure}
\end{figure}

\begin{figure*}[t!]
\centering
\includegraphics[width=0.97\textwidth]{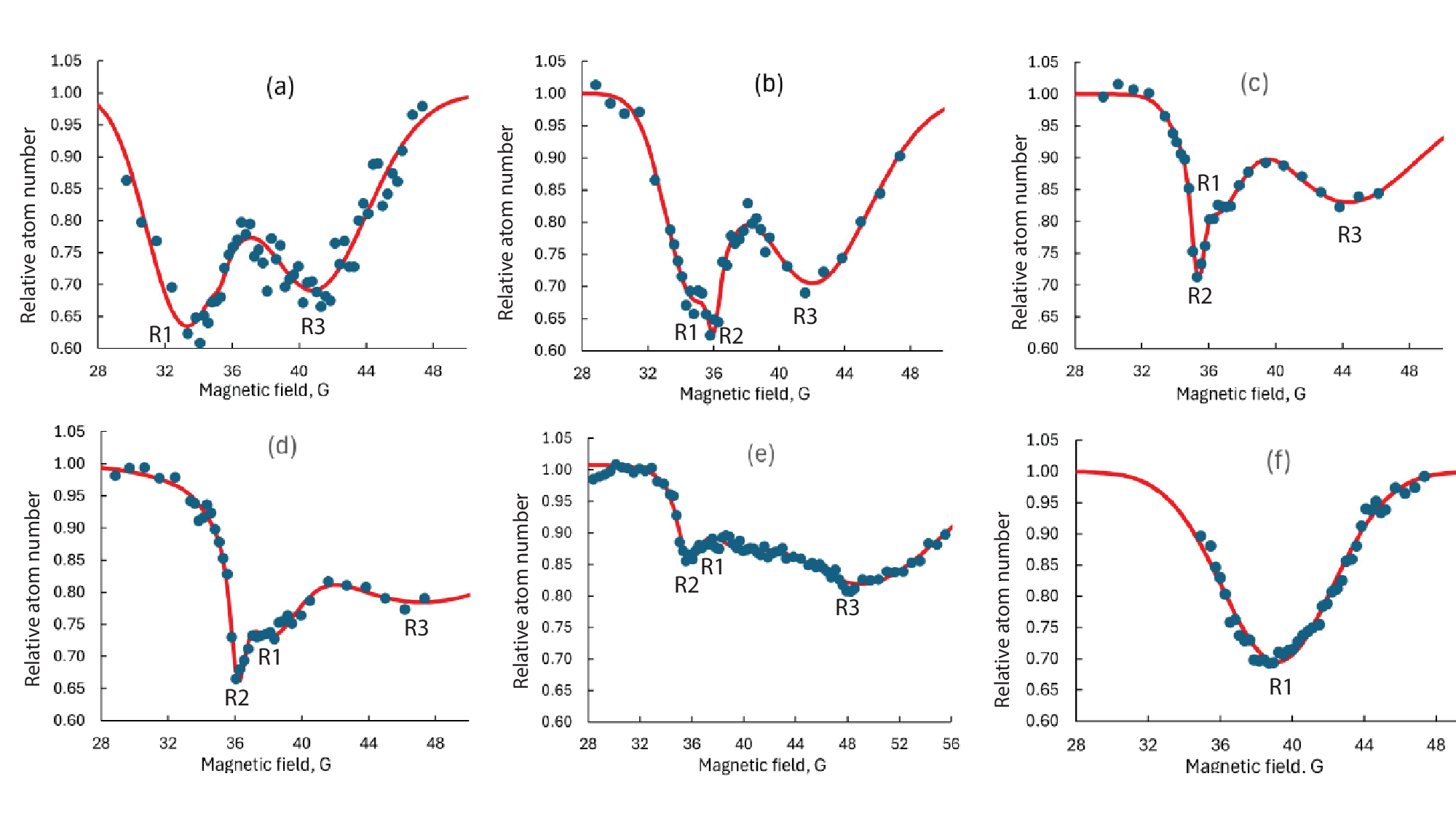}
\caption{(a)-(e) Scans of the R1, R2 and R3 Feshbach atom-loss resonances for an unpolarized-spin sample and a range of ODT powers and atom cloud temperatures: (a) 1.1 W, 6 $\mu$K; (b) 2.3 W, 10 $\mu$K; (c) 7.1 W, 20 $\mu$K; (d) 9.8 W, 25 $\mu$K; (e) 12 W, 30 $\mu$K. (f) Scan of the R1 resonance on a spin-polarized $|1, -1\ra$ sample, 15 W, 34 $\mu$K. Solid lines represent multi-Gaussian fits to the recorded profiles of the resonances.} 
\label{Figure}
\end{figure*}

\begin{figure*}[t!]
\centering
\includegraphics[width=0.99\textwidth]{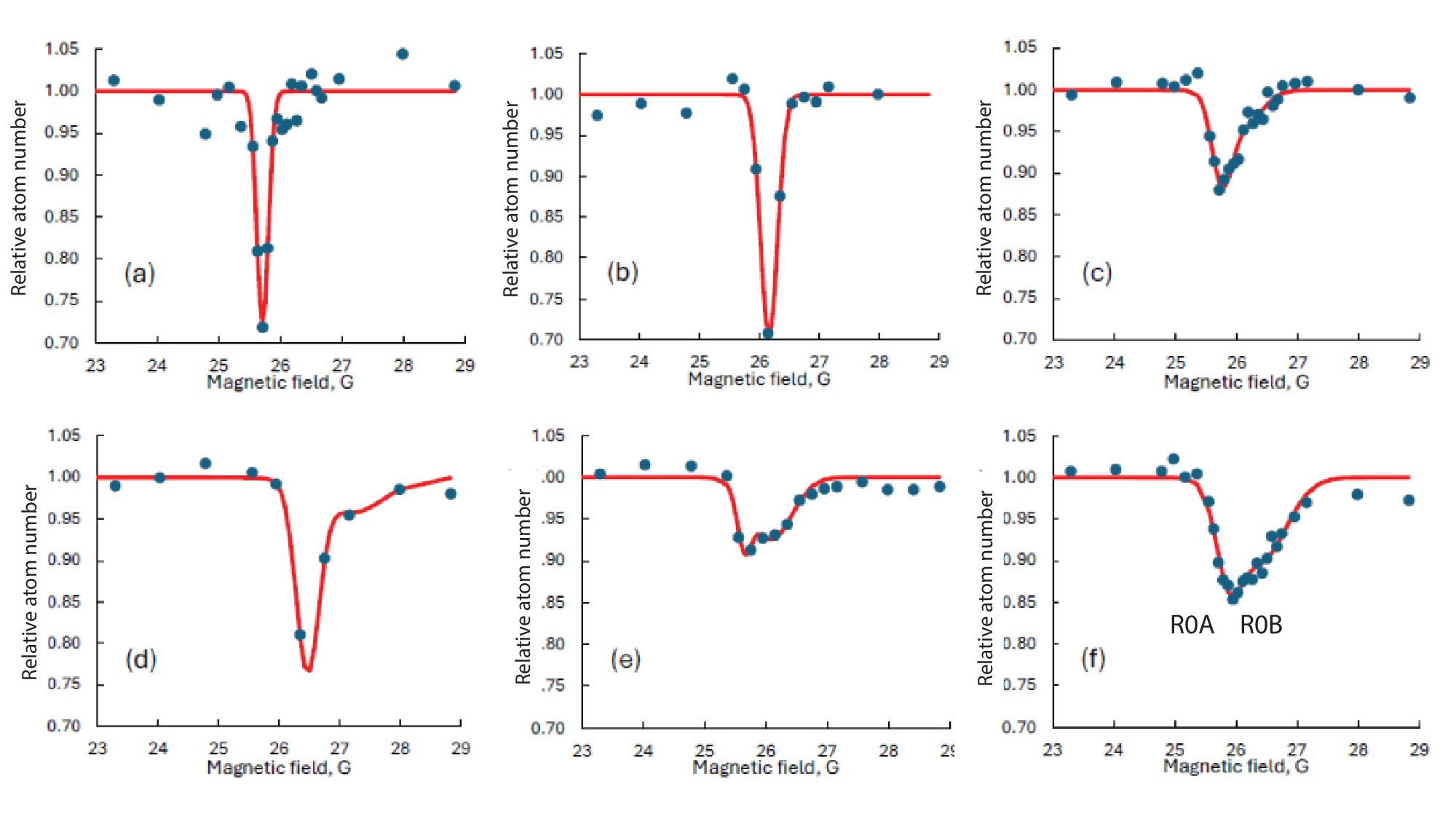}
\caption{Fine-step scans of the narrow 25.9 G (R0) Feshbach atom-loss resonance for a range of ODT laser powers and atom cloud temperatures, showing the evolution from a single resonance into a partially resolved doublet. (a) 1.1 W, 6.1 $\mu$K (b) 2.3 W, 9.4 $\mu$K, (c) 5.0 W, 15.5 $\mu$K, (d) 10 W, 25 $\mu$K (e) 12 W, 30 $\mu$K (f) 14 W, 34  $\mu$K. Solid lines represent single- or double-Gaussian fits to the profiles of the resonances.}
\label{Figure}
\end{figure*}

The number density and spatial profile of the trapped atoms are recorded by absorption imaging using a 767 nm $4S_{1/2}$, $F$=2 $\rightarrow$ $4P_{3/2}$, $F'$=3 imaging beam projected onto a 16-bit CCD camera (Princeton Instruments ProEm 512) operated in kinetics mode. The temperature of the atom cloud is determined from standard time-of-flight expansion measurements. 

The Feshbach magnetic field is controlled by an analog voltage signal from a National Instruments (NI) card that is fed into a Delta Elektronika (SM 18-220) power supply, leading to a magnetic field rms ripple of less than 100 mG. The magnetic field produced by the Feshbach coils was calibrated by recording the center positions of the R0, R1, R3, R5 and R6 resonances for a range of ODT laser intensities, and extrapolating to zero intensity and using the known positions of these resonances (Table I). A near-linear function with a small quadratic component was fitted to the magnetic field versus NI volts data to provide a calibration that was accurate to better than $1\%$.  

The Feshbach atom-loss resonances are recorded by applying the Feshbach magnetic field for 1 s at atom cloud temperatures ranging from 34 $\mu$K down to 6 $\mu$K by ramping down the depth of the ODT. For most of the experiments, the atom cloud in the crossed ODT was an unpolarized-spin sample, with the atoms occupying all three spin states, $|1, +1\ra$, $|1, 0\ra$ and $|1, -1\ra$, in the $4S_{1/2},F=1$ hyperfine multiplet. For some experiments on the 33.6 G resonance, the atom cloud was prepared in a predominantly spin-polarized $|1, -1\ra$ state by switching the Feshbach magnetic field to the 39.9 G $|1, +1\ra$  + $|1, 0\ra$ mixed-spin state resonance for typically 2 s to remove atoms in the $|1, +1\ra$ and $|1, 0\ra$ states. 

\section{Experimental Results}

\subsection{Assignment of $^{39}\textup{K}$ Feshbach Resonances}

Figure 3 shows an example scan of the Feshbach atom-loss resonances up to 65 G for an unpolarized-spin sample at a temperature of 16 $\mu$K and trap depth of 46 $\mu$K. 

The narrow resonance labeled R0 is the well-known 25.9 G resonance listed in Table I. The dip labeled R1 and R2, near 35 G, is a superposition of two resonances (Fig. 4): a broad resonance (1/e half-width $\approx$ 3.8 G), which we designate R1, and a narrow resonance ($\approx$ 0.5 G), which we designate R2. The broader resonance R1 shifts to lower magnetic fields with decreasing ODT laser power, approaching 33.4 G as the power approaches zero, and the amplitude increases from 16\% at 10 W ODT laser power up to 36\% at 1 W. This resonance is identified as the 33.6 G single-spin $|1, -1\ra$ resonance in the $\{cc\}$ channel. The position of the narrower resonance R2 remains fixed at 36 G for ODT laser powers from 14 W down to 1 W and the  amplitude decreases from 14\% at 14 W down to 2\% at 1 W. The two-body CC calculations (Fig. 1) indicate that R2 is neither a $s$-wave or a $d$-wave resonance and is possibly a $p$-wave spin-exchange resonance predicted in the $\{bc\}$ channel at 33 G. 

The resonance R3, near 43 G, which approaches 40.8 G as the ODT laser power approaches zero, is close to the 39.9 G resonance predicted in Fig. 1 and in previous calculations \cite{Lysebo10,Tiemann20,Hammond23}, and is identified from the CC calculations as a mixed-spin state $|1, +1\ra$ + $|1, 0\ra$ $\{ab\}$ resonance. We note that the observed R3 atom-loss resonance has a width comparable to that of the R1 resonance, whereas the width of the calculated R3 is much narrower than R1 (Fig. 1). The larger width of the R3 resonance may possibly result from large inelastic collision losses. The resonance R4, near 53 G, corresponds closely to the 52.6 G $p$-wave spin-exchange resonance predicted in the $\{bc\}$ channel by the CC-calculations. The resonances R5 and R6, near 59 G and 64 G, are the well-known resonances in the $\{bb\}$ channel (Figs. 1 and 2). 

In Fig. 4, we present scans of the R1, R2 and R3 Feshbach atom-loss resonances for a range of  ODT laser powers.  With decreasing ODT laser power, the R1 33.6 G resonance moves across the fixed R2 36.0 G resonance, confirming that R1 shifts significantly with ODT laser power. Also shown, in Fig. 4(f), is a scan of the R1 resonance taken on a spin-polarized $|1, -1\ra$ sample, which exhibits a well-resolved single resonance without any traces of the mixed-spin R2 and R3 resonances. The position of this single resonance approaches 33.4 G as the ODT laser power or trap depth approaches zero, confirming the assignment of the R1 resonance in Figs. 3 and 4. Moreover, since the spin-polarized $|1, -1\ra$ sample was prepared using the 39.9 G  $|1, +1\ra$ +  $|1, 0\ra$ mixed-spin state resonance to remove atoms in the  $|1, +1\ra$ and $|1, 0\ra$ states, the observation of a single R1 resonance  provides additional confirmation that the R3 resonance is a $|1, +1\ra$ +  $|1, 0\ra$ mixed-spin state resonance.

\begin{figure}[t]
\centering
\includegraphics[width=0.62\textwidth]{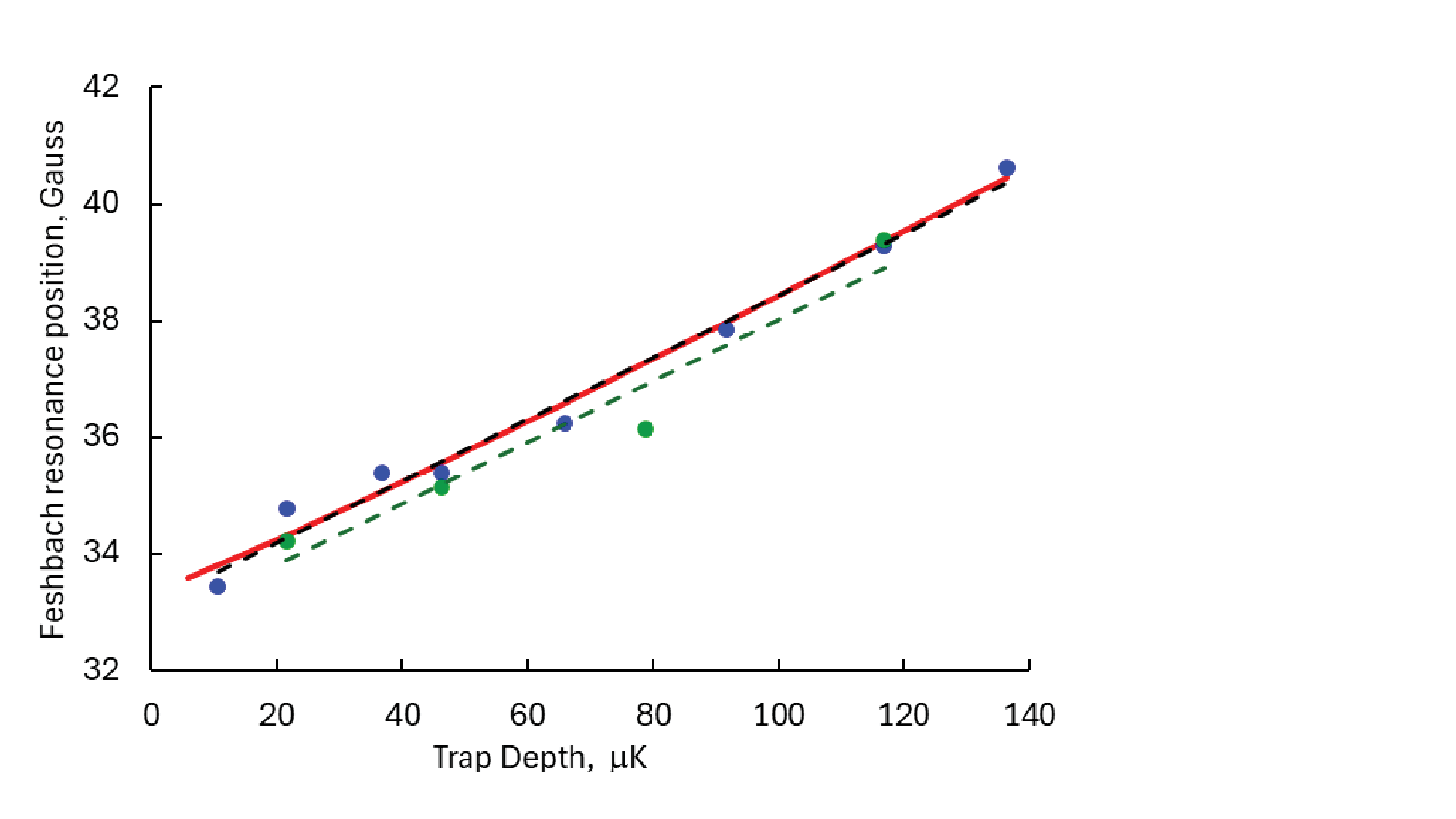}
\caption{Measured center position of the R1 33.6 G Feshbach resonance versus ODT trap depth (in $\mu$K). Blue points: temperature allowed to vary for a $^{39}\textup{K}$ unpolarized spin  sample; green  points: temperature kept approximately constant at an average value of 16(2) $\mu$K for a spin-polarized  $|1, -1\ra$ sample. The red solid line represents a fit to the spin-unpolarized data using Eq (1) and the black dashed line represents a straight-line fit to the same data. The green dashed line represents a straight-line fit to the spin-polarized data.} 
\label{Figure}
\end{figure}

Figure 5 shows fine-step scans of the narrow R0 resonance at 25.9 G for a range of ODT laser powers.  With increasing ODT laser power, the atom-loss resonance evolves from a single resonance into a partially resolved doublet. According to the CC-calculations in Fig. 1 and previous Feshbach resonance calculations \cite{Lysebo10, Hammond23}, the R0 resonance is predicted to consist of two near-coincident resonances: a single-spin state resonance (R0A) in the $\{aa\}$ channel associated with the (1, 3, 3) molecular state  and a mixed-spin state resonance (R0B) in the $\{ab\}$ channel associated with the (1, 1, 1,) molecular state (Fig. 2), with R0B about three-times broader than R0A (Fig. 1 and \cite{Lysebo10}). To the best of our knowledge the predicted ROB resonance has not been observed before.

\subsection{Shifts of the Feshbach Resonances}

In Fig. 6, we present measurements of the center position of the R1 33.6 G Feshbach resonance for an unpolarized-spin sample (blue points) as the atom cloud is forced evaporatively cooled to lower temperatures by reducing the power of the 1063.9 nm ODT laser and hence the trap depth of the ODT. The position of the resonance decreases almost linearly with decreasing trap depth, with a zero trap-depth intercept of 33.4(2) G, which is close to the known position of the 33.6 G Feshbach resonance (Table I). 
\begin{figure}[t]
\centering
\hspace*{-0.5 cm}
\includegraphics[width=1.06\textwidth]{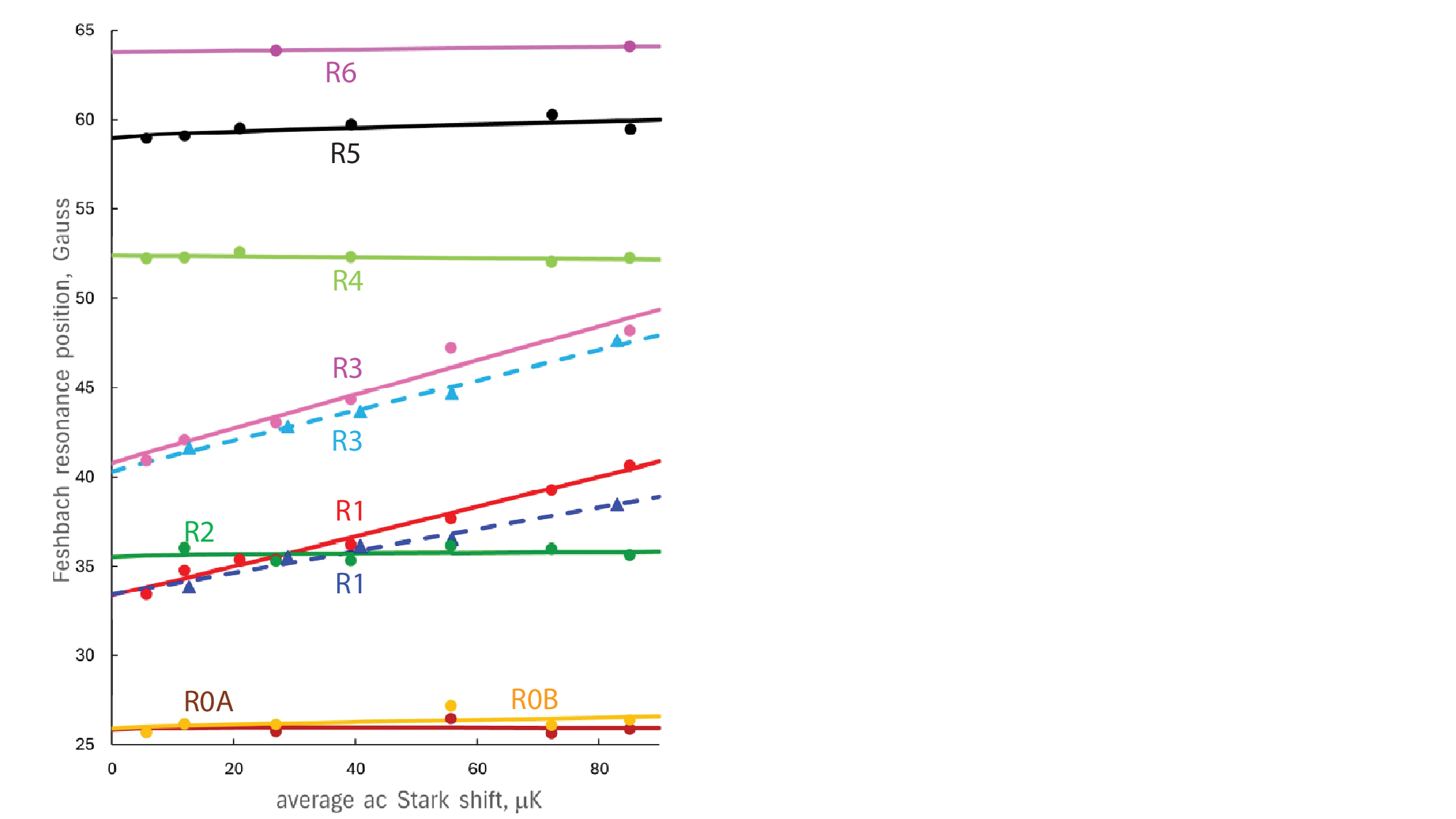}
\caption{Dependence of measured center positions of the studied $^{39}\textup{K}$ Feshbach resonances on trap-averaged ac Stark shift (in $\mu$K) and an ODT laser frequency of 281.7665(2) THz for R1-R6 (circles) and 281.7796(2) THz (triangles) for R1 and R3. The lines represent fits to the data points using Eq (1), with the dashed lines for R1 and R3 referring to the ODT laser frequency of 281.7796(2) THz. The labels R0A and R0B refer to the resonances at 25.9 G, R1 to 33.6 G, R2 to 36 G, R3 to 39.9 G, R4 to 52.6 G, R5 to 58.9 G, and R6 to 65.6 G. } 
\label{Figure}
\end{figure}

Two effects can shift the position of a Feshbach resonance. First, the finite temperature of the trapped atoms leads to increased collision energy, resulting in a small shift of the resonance, usually towards higher collision energy \cite{Khlebnikov21}. The observed shift of about +7.5 G at 35 $\mu$K is more than an order of magnitude larger than, and of opposite sign to, the estimated temperature-induced shift of about $-0.009$ G/$\mu$K based on a simple $\frac{3}{2}k_{B}T$ thermal energy dependence and a differential magnetic moment of $-2.5 \mu_B$ (Table I). The temperature shift is towards low magnetic fields for the R1 resonance because of the negative differential magnetic moment of the associated molecular state at the magnetic field of the Feshbach resonance. 

Secondly, the oscillating electric field of the 1063.9 nm ODT laser beams induces a differential ac Stark shift between the incoming pair of $^{39}\textup{K}$ atoms and the quasi-bound state of the associated Feshbach molecule \cite{Bauer09, Vexiau11, Kohstall12, Cetina15, Gregory17, Lous18, Peng18, Khlebnikov21, Spence23, Wei26}, resulting in a shift in the position of the resonance given by (see Appendix):

\begin{equation}
\delta \mu\Delta B = \frac{\delta\alpha}{\alpha_{at}} U_{av}(U_0, T) + \Delta U_{th}, 
\end{equation}

\noindent where $\delta \mu = \mu_{oc} - \mu_{cc}$ is the difference between the magnetic moment of the open (atomic) channel $oc$ and the closed (molecular)
channel $cc$, $\delta\alpha = 2 \alpha_{at}-\alpha_{mol}$ is the differential polarizability between the atom-pair and the Feshbach molecule, $U_{av}(U_0, T)$ is the spatially averaged trap depth, and $\Delta U_{th}$ is the difference in thermal energies of the two atoms and the associated molecule. 

\begin{table*}[ht]
\begin{center}
\caption{Measured shifts of the center positions of the $^{39}\textup{K}$ Feshbach resonances (in mG/${\mu}$K (Stark)) and extracted polarizabilities of the molecules $\alpha_{mol}/\alpha_{at}$ associated with the Feshbach resonances, using the updated $\delta \mu = \mu_{oc} - \mu_{cc}$ values in Table I. The results in columns 4 and 5 were obtained for an ODT laser frequency of 281.7665 THz and those in columns 6 and 7 for an ODT laser frequency of 281.7796 THz. The polarizability of the potassium atom at 1063.9 nm (281.8 THz) is 598.7 a.u. \cite{Safronova13, Ravensbergen18, kiruga26}. Uncertainties represent one standard error. [tw] refers to this work.}
\vspace{2mm}

\begin{tabularx}{\textwidth}{|c|c|c|c|c|c|c|}
\hline
\hline
\@ \@ \@ \@ Label\@ \@ \@ \@ \@ & \@ \@ \@ $B_0$ (G) \@ \@ \@ \@ \@  & $B_0$ (G) @ $I$=0  & \@ Shift (mG/${\mu}$K) \@   \@ \@ & $\alpha_{mol}$/ \@ $\alpha_{at}$ &\@ \@ \@ Shift (mG/${\mu}$K) \@ & \@ $\alpha_{mol}$/$\alpha_{at}$ \@\\
& \@ \@ \@ \@ \@ \@ Theory \@ \@ \@ \@ \@ \@ \@ \@   & \@ Experiment [tw] \@ & \@ \@ @281.7665 THz \@ \@ & @281.7665 THz \@ \@ & @281.7796 THz & \@ @281.7796 THz \@ \\  

\hline
R0A & 25.9 \cite{Derrico07} & 25.9(3) & 0(5) & +1.5(5) & +3(2) & +1.8(2) \\
\hline
R0B & \@ 25.8 \cite{Lysebo10} & 25.9(4) & +7(7) &  +2.2(7) & & \\
\hline
R1 & 33.6 \cite{Derrico07} & 33.4(2) &  +83(5) & $-12.5(1.6)$ & +61(7) & $-8.8(1.1)$  \\
\hline
R2 & & 35.5(3) & +2(7) & & +10(4) & \\
\hline
R3 & \@ \@ 39.86 \cite{Lysebo10} & 40.7(5) & +95(11) & +9(1) & +84(6) & +8.1(5) \\ 
\hline
R4 & \@ 52.6 [tw] & 52.4(1) & $-3(2)$  &  & $-14(29)$ &  \\
\hline
R5 & 58.8 \cite{Derrico07} & 59.0(3) & +10(5) & +2.0(2) &  & \\
\hline
R6 & 65.6 \cite{Derrico07} & \ 65.0(7) & +4 & +1.8 & $-19(15)$ & 0.0(1.0)\\
\hline
\end{tabularx}
\end{center}
\end{table*}

During forced evaporative cooling, the temperature of the atom cloud is normally proportional to the depth of the ODT (Fig. 10 in Appendix). We can vary the ODT trap depth while keeping the temperature of the atom cloud approximately constant by slowly ramping down the intensity of the ODT beam in about 400 ms to first cool the atom cloud to a certain temperature and then rapidly switching the intensity of the ODT beam back to the initial value in about 1 ms to a required temperature followed by an equilibration time of 1000 ms. The green points in Fig. 6 show the position of the R1 33.6 G resonance versus ODT trap depth with the temperature of the atom cloud kept fixed at an average value of 16(2) $\mu$K, apart from the lowest trap-depth point which was recorded at 8 $\mu$K. The slope [0.053(9) G/$\mu$K (trap depth)] and zero trap-depth intercept [32.7(7) G)] of the straight-line fit to the green points are essentially the same as the slope [(0.053(3) G/$\mu$K (trap depth)] and intercept [33.1(2) G] when the temperature is allowed to vary (blue points). This indicates that the observed large shifts in the position of the 33.6 G resonance occur predominantly as a result of the change in trap depth of the ODT, rather than by the change in temperature of the atom cloud during the lowering of the ODT power. 

In an ODT, the atoms and the associated Feshbach molecules experience an ac Stark shift depending on how far they climb the ODT walls. We measure the remaining atom number in the trap, so the precise measurable is determined not by the trap depth but by the ac Stark shift experienced by the whole ensemble (as described in the Appendix). Figure 7 shows the dependence of the measured Feshbach center position on the trap-averaged ac Stark shift (in $\mu$K) for the eight studied Feshbach resonances and ODT laser frequencies of 281.7665(2) THz (circles), and 281.7796(2) THz (triangles) for R1 and R3. For the ODT laser frequency of 281.7665(2) THz, the slopes of the fits of Eq (1) to the data points are large for R1, +83(4) mG/$\mu$K (Stark), and R3, +95(11) mG/$\mu$K (Stark), and essentially zero for the other six resonances (Table II). The slope of the fit for R1 in Fig. 7 differs from that in Fig. 6 because of the different abscissae. The intercepts of the fitted curves are close to the known center positions of the various Feshbach resonances (Table II). For the higher ODT laser frequency of 281.7796(2) THz, the slopes of the fits again reveal large shifts for the R1 and R3 resonances and esentially zero shifts for the other resonances, but at this ODT laser frequency, the shift for R1, +61(7) mG/$\mu$K (Stark), is significantly smaller, while the shift for R3, +84(6) mG/$\mu$K (Stark), is only slightly smaller. 

We use the measured shifts of the center positions of the Feshbach resonances, $\Delta B = B_1 - B_0$,  as a function of the trap-averaged ac Stark shift to extract the real part of the dynamic polarizability of the associated Feshbach molecules at the frequency of the ODT laser. The procedure compares the molecular polarizability with the known polarizability of the potassium atom at 1063.9 nm (598.7 a.u. \cite{Safronova13, Ravensbergen18, kiruga26}), using Eq (5) of the Appendix, similar to the procedure previously employed to determine the polarizability of dysprosium atoms relative to potassium atoms as a reference \cite{Ravensbergen18}.  

The polarizabilities of the Feshbach molecules determined from the measured Feshbach shifts versus ac Stark shift of the resonances are summarized in Table II. At the 281.7665 THz ODT laser frequency, the molecular polarizabilities associated with the 33.6 G and 39.9 G resonances are $\alpha_{mol} = -12.5(1.6)\alpha_{at}$ and $+9(1)\alpha_{at}$, respectively, while at the higher 281.7796 THz ODT laser frequency, the polarizabilities are $\alpha_{mol} = -8.8(1.1)\alpha_{at}$ and $+8.1(5)\alpha_{at}$, respectively. The negative sign of the polarizability in the case of the 33.6 G resonance, is opposite to the sign of the slope of the position shift versus ac Stark shift curve (Fig. 7) owing to the negative differential magnetic moment, $\delta \mu = -2.5\mu_B$, at the Feshbach magnetic field, see Sec. II and Table I.  

\section{Discussion}

We attribute the large dynamic polarizabilities of the Feshbach molecules associated with the 33.6 G and 39.9 G resonances to a near-coincidence between the frequency 281.7665 THz of the 1063.9 nm ODT laser and the frequency of a $\textup{K}_2$ molecular transition from the last vibrational level $v''=26$\ \cite{Pashov08} within the metastable $a{^3\Sigma_u^+}$ potential to a vibrational level within an excited triplet potential. This interpretation is supported by the observation that the magnitude of the shifts of the 33.6 G and 39.9 G resonances depends on the frequency of the ODT laser (Table II). The only $\textup{K}_2$ triplet transitions from the $a{^3\Sigma_u^+}$ vibrational levels at wavelengths near 1063.9 nm are the lowest triplet transitions $a{^3\Sigma_u^+}(v'') \rightarrow \, b{^3\Sigma_g^+}(v')$, which have been studied experimentally in high-density gases in \cite{Huennekens84, Vadla06} and theoretically in \cite{Vadla06, Beuc14}. 

Interaction potentials and vibrational-level data are well established for the lower $a{^3\Sigma_u^+}$  (e.g., \cite{Pashov08}), but only limited data are available for the upper $b^3\Sigma_g^+$ \cite{Ligare91,Magnier96, Magnier04,Vadla06, Beuc14}. The $b{^3\Sigma_g^+}$ potential reported in \cite{Vadla06,Beuc14}, which is based on unpublished data \cite{Yan}, has a dissociation limit, $D_e \approx 6,300\  \textup{cm}^{-1}$, which is much larger than the $D_e = 3,786\   \textup{cm}^{-1}$ reported in the \textit{ab-initio} calculations of Magnier et al. \cite{Magnier04}, and it also appears to be inconsistent with the observed position of the $a{^3\Sigma_u^+}$ bandhead in the experimental $a{^3\Sigma_u^+}(v'') \rightarrow \, b{^3\Sigma_g^+}(v')$ spectra in \cite{Huennekens84} and \cite{Vadla06}. In the following, we adopt the $b^3\Sigma_g^+$ potential provided by the Orsay group \cite{Vexiau26}, which is in agreement with calculations of the spectroscopic constants of Magnier et al. \cite{Magnier04}. Figure 8 shows the interaction potentials for the metastable $a{^3\Sigma_u^+}$ \cite {Pashov08} and the excited $b{^3 \Sigma_g^+}$ \cite{Vexiau26}, along with the ground  $X{^1\Sigma_g^+}$ \cite {Pashov08}.

In the absorption spectrum of the $a{^3\Sigma_u^+}(v'') \rightarrow \, b{^3\Sigma_g^+}(v')$ transitions, the peak wavelength of the narrow $a{^3\Sigma_u^+}$ bandhead has been observed at 1095 nm \cite{Huennekens84} and 1096 nm \cite{Vadla06}, or about 2.10 THz below the 281.766 THz frequency of the ODT laser. This indicates that at the ODT laser frequency there is significant Franck-Condon overlap of the upper and lower vibrational-level wavefunctions which, along with the vibrational level-spacing of about 1.77 THz \cite{Magnier04} near the bottom of the potential well, indicates that the frequency of the ODT laser is close to the transition $a{^3\Sigma_u^+ (v''=26)} \rightarrow \, b{^3\Sigma_g^+ (v'=4)}$. The specific 1063.9 nm transitions associated with the large observed polarizabilities for the 33.6 G and 39.9 G resonances are likely to be transitions from the $(S, I, f, M)=(1, 1, 2, -2)$ and (1, 3, 3, +1) molecular hyperfine Zeeman states, respectively, within the last rovibrational level $a{^3\Sigma_u^+}(v''=26\ , N=0)$ to hyperfine Zeeman states within the  rovibrational level $b^3\Sigma_g^+v'=4, N=1)$.

\begin{figure}[t]
\centering
\hspace*{0.0 cm}
\includegraphics[width=0.91\textwidth]{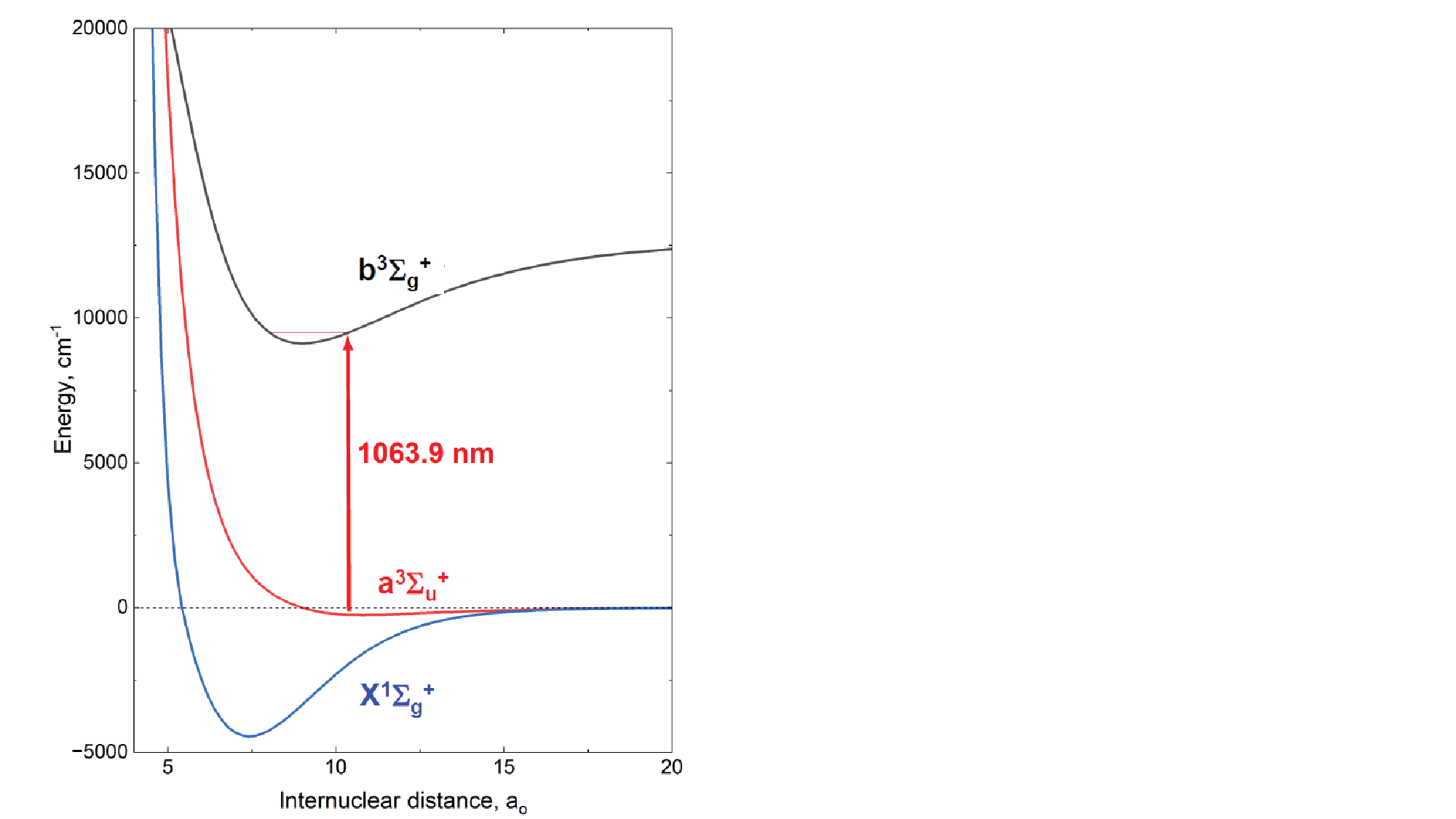}
\caption{Calculated interaction potentials for the triplet $a{^3\Sigma_u^+} \rightarrow \, b{^3\Sigma_g^+}$ transition at 1063.9 nm in the $\textup{K}_2$ molecule based on the data of \cite{Pashov08} for the metastable $a{^3\Sigma_u^+}$ and \cite{Vexiau26} for the excited $b{^3 \Sigma_g^+}$ potential. Also shown is the $X{^1 \Sigma_g^+}$ ground potential \cite{Pashov08}.} 
\label{Figure}
\end{figure}

The molecular polarizability $\alpha$ can be defined in terms of a parallel component $\alpha_\parallel$ along the molecular axis, involving $^3 \Sigma_u ^+ \rightarrow \, ^3 \Sigma_g ^+$ transitions, and a perpendicular component $\alpha_\perp$, involving $^3 \Sigma_u^+  \rightarrow \,^3\Pi_g $ transitions \cite{Deiss15,Deiss14}. The ODT laser frequency is close to the frequency of the $a{^3\Sigma_u^+} \rightarrow$ \,$b{^3\Sigma_g^+}$ transitions at 1063.9 nm but is much smaller than the frequency of  the $a{^3\Sigma_u^+} \rightarrow \, b{^3\Pi_g^+}$ transitions at around 721 nm. This implies that the observed polarizabilities are $\alpha_\|$, i.e., parallel to the molecular axis.  

Assuming the polarizability associated with the 33.6 G resonance has a single dispersion-shaped frequency distribution with natural linewidth two-times the atomic linewidth of 6 MHz, the observed change of $\alpha_{mol}/ \alpha_{at} = {-}12.5(1.6)$ to $-8.8(1.1)$ when the ODT laser frequency is shifted by +13.1 GHz is consistent with the two ODT laser frequencies having detunings of about +40 GHz and about +50 GHz, respectively. The negative and positive polarizabilities of the Feshbach molecules associated with the 33.6 G and 39.9 G resonances suggest that the ODT laser frequency occurs on the blue-detuned side of the dispersion-shaped polarizability distribution for the 33.6 G resonance and on the red-detuned side for the 39.9 G resonance. The polarizability of the Feshbach molecule associated with the 39.9 G resonance would then be expected to increase with increasing ODT laser frequency; however, we observe a small (barely significant) decrease from $\alpha_{mol}/\alpha_{at} = +9(1)$ to $+8.1(5)$ ($1\sigma$ error) for the 13.1 GHz frequency increase. This suggests that the frequency distribution of the polarizability may contain some structure, which could arise, for example, from residual oscillations similar to those found in the central absorption region in $\textup{Cs}_2$ \cite{Vexiau11} and $\textup{Rb}_2$ \cite{Deiss15}. 

We  note that, of the eight resonances studied, the Feshbach molecule associated with the 33.6 G resonance is the only one that has a negative  polarizability. It is also the only resonance that is associated with the molecular hyperfine state $(S, I, f)$=(1, 1, 2), while the other resonances are associated with the (1, 3, 3) or (1, 1, 1) hyperfine states, which are degenerate at zero magnetic field (Sec. II) and separated from the (1, 1, 2, -2) hyperfine Zeeman state by about 80 MHz at the Feshbach magnetic fields (Fig. 2).  

There remains the question of why the large anomalous shifts occur for the 33.6 G and 39.9 G resonances and not the other $^{39}\textup{K}$ resonances. The differences in detunings of the transition frequencies of the various Feshbach molecules from the ODT laser frequency are too small ($\approx 80$ MHz) compared with the estimated detunings ($\approx 50$ GHz) and with the vibrational level-spacing of the upper $b{^3\Sigma_g^+}$ ($\approx 1770$ GHz) to account for the large difference in position shifts of the 33.6 G and 39.9 G resonances. Also, it seems unlikely that the different quantum numbers $(S, I, f, M)$ involved in the various molecular transitions can affect the transition rates of the induced molecular transitions in  a way consistent with the observed differences in the position shifts. A distinguishing characteristic of the 33.6 G resonance is that it is a broad intermediate-strength ($s_{res} = 1.9$) resonance with substantial contributions from both the closed-channel bound state and the open-channel free-atom state, while the other narrow resonances are closed-channel dominated resonances (Table I). However, the ac Stark shifts of the molecular states are expected to occur during the closed-channel (molecule) phase rather than during the open-channel (free-atom) phase, and so the position shifts are not expected to be sensitive to the coupling strength. Clearly, a full theoretical treatment is required to quantitatively account for the observed position shifts.

In many studies of Feshbach molecules (e.g., \cite{Danzl10, Vexiau11, Deiss14, Deiss15, Vexiau17}), the polarizability of the weakly bound Feshbach molecule is taken to be the sum of the polarizabilities of the two incoming atoms. This is also the case for six of the $^{39}\textup{K}$ Feshbach molecules we have studied. The strongly enhanced polarizabilities for the Feshbach molecules associated with the $^{39}\textup{K}$ 33.6 G and 39.9 G resonances represent major exceptions to this general behavior.   

It is reasonable to ask if the large position shifts for the 33.6 G and 39.9 G Feshbach resonances are likely to occur in $^{39}\textup{K}$ experiments in other laboratories that use a nominal 1064 nm ODT fiber laser? Assuming a $1/\Delta \omega$ fall-off of the polarizability with detuning $\Delta \omega$, we estimate that to achieve a 1 G shift of the 33.6 G resonance in an ODT with a trap depth of 140 $\mu$K, the detuning needs to be less than about 300 GHz, which is larger than the typical tolerance of $\pm$130 GHz ($\pm$0.5 nm) for a 1064 nm single-frequency fiber laser \cite{Azur26}.    

In future studies, it would be interesting to map the entire frequency distribution of the polarizability of the Feshbach molecules associated with the 33.6 G and 39.9 G resonances and also the other $^{39}\textup{K}$ resonances in Table I using, for example, a tuneable single-frequency fiber laser. This could form the basis of a new molecular spectroscopy that would provide unique information on the frequency distribution of the polarizability of Feshbach molecules. It would also be interesting to see if large shifts of Feshbach resonances and large positive and negative polarizabilities occur for Feshbach molecules in other atomic systems.

\section{Summary and Conclusions}

Anomalous large shifts of the positions of the 33.6 G and 39.9 G Feshbach resonances have been observed for $^{39}\textup{K}$ atoms confined in a 1063.9 nm optical dipole trap at temperatures up to about 35 $\mu$K and trap depths up to about 136 $\mu$K. The large shifts originate from large differential ac Stark shifts between the incoming pair of $^{39}\textup{K}$ atoms and the associated quasi-bound Feshbach molecule in the oscillating electric field of the optical dipole trap. The large differential ac Stark shifts in turn originate from an unusually large negative dynamic polarizability of the weakly-bound Feshbach molecule in the case of the 33.6 G resonance and from a large positive polarizability in the case of the 39.9 G resonance. 

The magnitude of the polarizabilities of the molecular states associated with the 33.6 G and 39.9 G Feshbach resonances are about four to seven times higher than the molecular polarizabilities for the other studied $^{39}\textup{K}$ Feshbach resonances, which are close to the sum of the polarizabilities of the two incoming $^{39}\textup{K}$ atoms. The large molecular polarizabilities associated with the 33.6 G and 39.9 G resonances are attributed to a near-coincidence (within $\approx$ 50 GHz) between the frequency of the optical dipole trap laser and the frequency of a transition from a molecular hyperfine state within the last rovibrational level of the metastable $a{^3\Sigma_u^+}$ to an excited state within a rovibrational level of the $b{^3\Sigma_g^+}$ of the $\textup{K}_2$ molecule.

\section*{Appendix}

\subsection{Relationship between the shift of a Feshbach resonance and the dynamic molecular polarizability}

The measured shift of the center position of a Feshbach resonance arising from a differential ac Stark shift allows a determination of the real part of the dynamic polarizability of the associated Feshbach molecule, using the known polarizability of the  $^{39}\textup{K}$ atom \cite{Ravensbergen18} as a reference.

\begin{figure}[t]
\centering
\hspace*{0.0cm}
\includegraphics[width=0.65\textwidth]{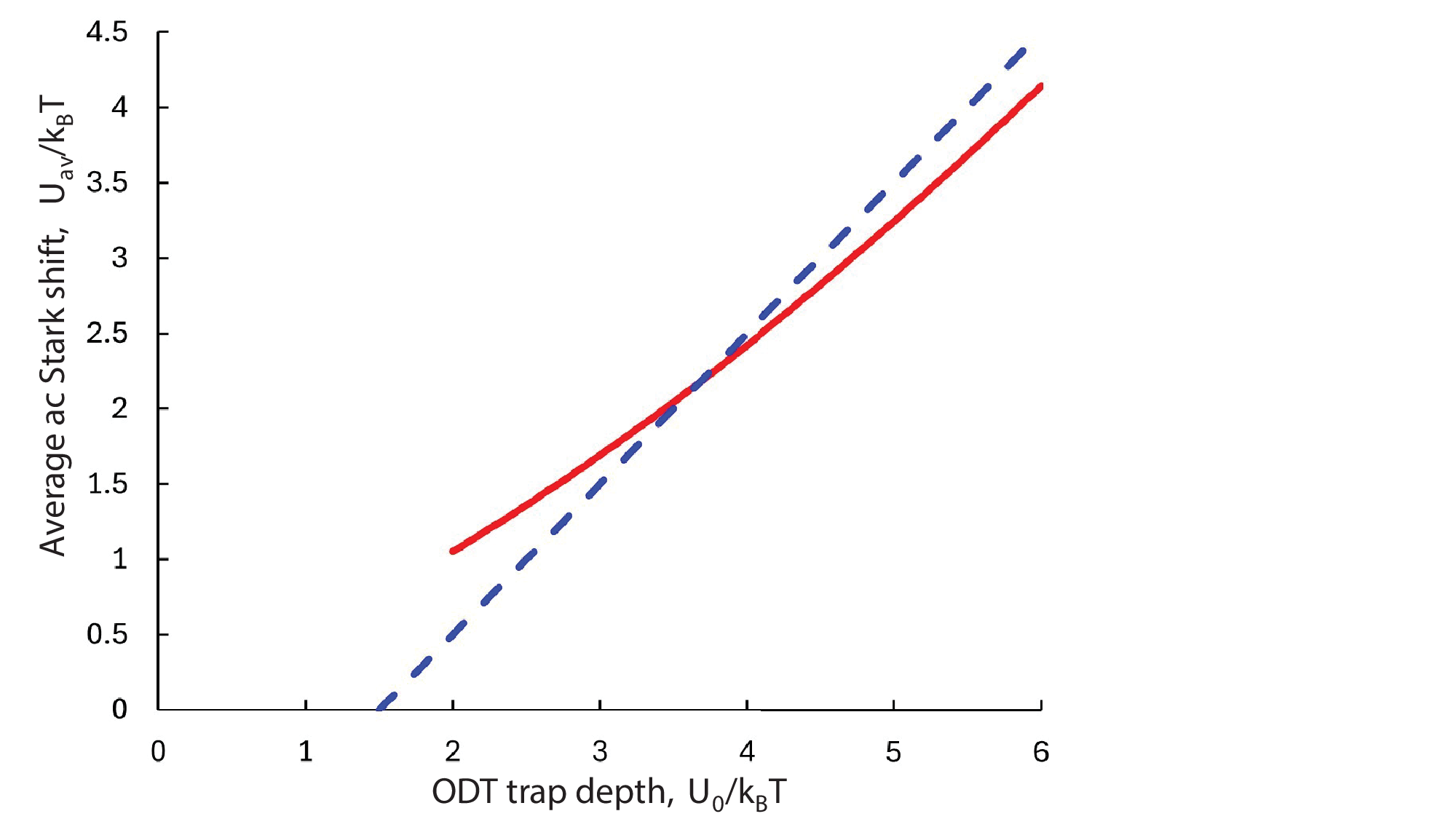}
\caption{Calculated spatially averaged optical dipole trap potential (ac Stark shift) as a function of the trap depth, assuming a 3D-Gaussian trapping potential and a truncated Boltzmann energy distribution (red curve). The dashed blue line represents a simple model $U_{av}=U_{0} -\frac{3}{2}k_{B}T$, as described in the text.} 
\label{Figure}
\end{figure}

We consider a crossed-beam optical dipole trap (ODT) formed by two Gaussian beams having orthogonal polarizations. The trapping potential, or the spatially varying ac Stark shift of the atomic state at the laser frequency $\omega_L$, in the oscillating electric field of the ODT laser is

\begin{equation}
U_{at}(\mathbf{r}, \omega_L)=-\frac{2\pi a_0^3}{c} \alpha_{at}(\omega_L) I(\mathbf{r}),
\end{equation}

\hspace{4mm}

\noindent where $\alpha_{at}(\omega_L)$ is the polarizability of the atomic state in atomic units, $I(\mathbf{r})$ is the spatially varying light intensity in the crossed-beam ODT, $a_0$ is the Bohr radius, and $c$ is the speed of light. In an ODT potential, the shift in the center position  of a Feshbach resonance, $\Delta B=B_1-B_0$, arising from a differential ac Stark shift between the molecular and atomic-pair potentials is related to the difference of the real part of the dynamic polarizabilities of the atom pair and the molecule $\delta\alpha = 2\alpha_{at}-\alpha_{mol}$: 

\begin{equation}
\delta \mu \Delta B(\mathbf{r},\omega_L) =  \frac{\delta\alpha}{\alpha_{at}} U_{at}(\mathbf{r},\omega_L) + \Delta U_{th},
\end{equation}

\noindent where $\delta \mu = \mu_{oc} - \mu_{cc}$ is the difference between the magnetic moment of the open (atomic) channel $oc$ and the closed (molecular) channel $cc$, and $\Delta U_{th}$ is the difference in thermal energies of the two atoms and the associated molecule. 

We record the total atom number in the crossed-beam region; so atomic-ensemble-averaging of the trapping potential is required. In our analysis, we assume a spherically symmetric 3D-Gaussian trapping potential and a truncated Boltzmann energy distribution for the atom cloud with forced evaporation. The atom number distribution can then be expressed as

\begin{equation} 
n(r) = n_0 \exp(-\tilde{u}_{at}) \times \frac{2}{\sqrt{\pi}} \gamma_{3/2}(-\tilde{u}_{at}),
\end{equation}

\noindent where $\tilde{u}_{at}=-U_{at}(\mathbf{r},\omega_L)/(k_B T)$, 
$\gamma_{3/2}(x)=\int_{0}^{x} t^{1/2} e^{-t} dt$ is the lower incomplete gamma function characterizing the truncation of the Boltzmann energy distribution, $n_0=\frac{\sqrt\pi}{2}\Lambda^{-3}\exp(\frac{\mu_0}{k_B T})$ is the peak number density $-$ determined by the thermal de Broglie wavelength $\Lambda$ and the chemical potential $\mu_0$, $k_B$ is Boltzmann's constant, and $T$ is the temperature of the trapped atoms. We perform numerical simulations of the dependence of the averaged ac Stark shift
$U_{av}=-4\pi\int_{0}^{\infty}dr\cdot r^2\cdot\frac{n(r)}{N}\cdot U_{at}(r,\omega_L)$ on the ODT bottom $U_{0} =(2\pi a_0^3/c) \alpha_{at} I_0$, where $I_0$ is the peak intensity, with results presented in Fig. 9 (red curve). The dashed blue line in Fig. 9 represents a simple model
$U_{av}=U_0-\frac{3}{2}k_B T$ based on a 3D harmonic potential and $\frac{1}{2}k_B T$ of potential energy per degree of freedom.

\begin{figure}[t]
\centering
\hspace*{0.0cm}
\includegraphics[width=0.92\textwidth]{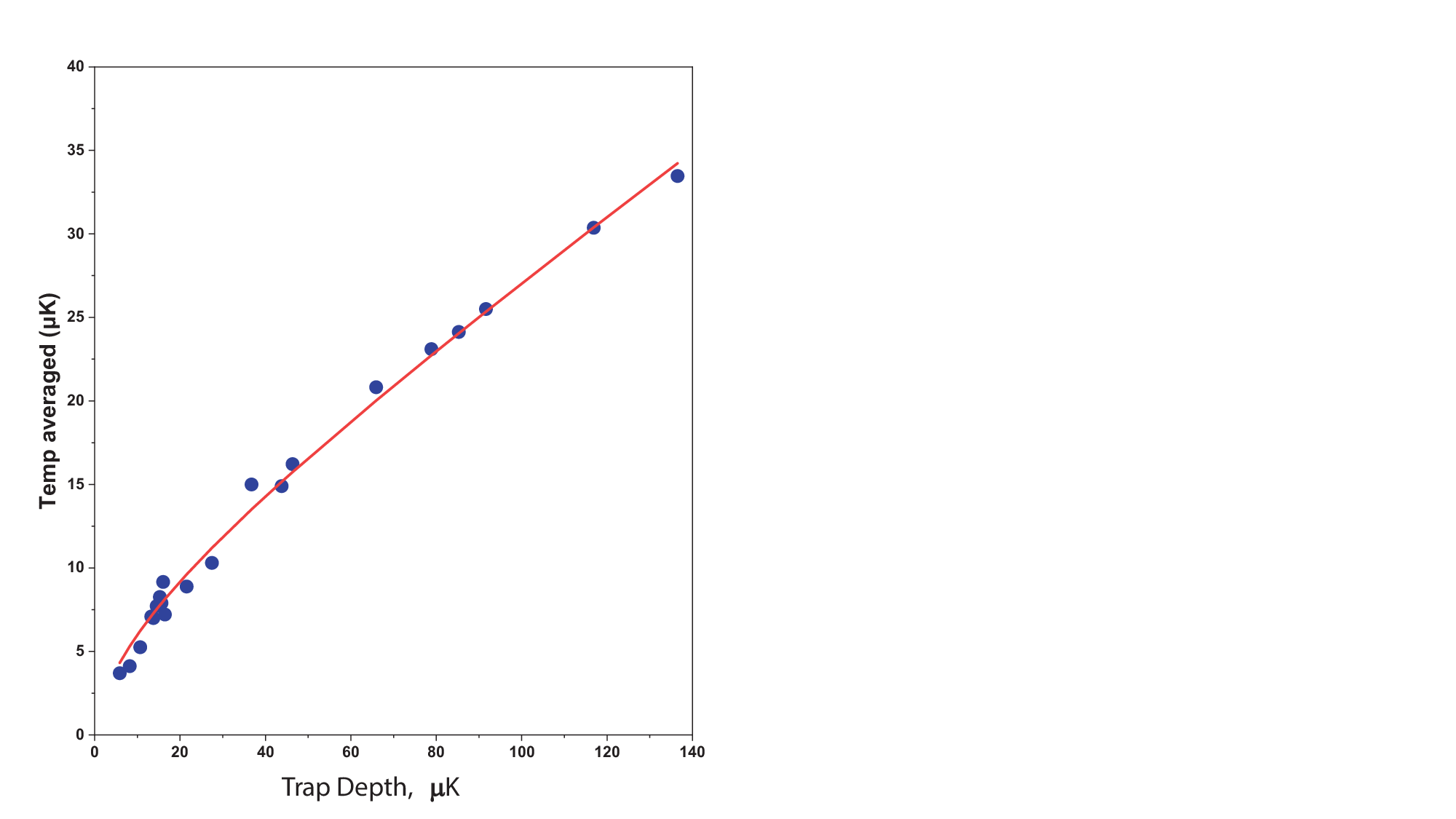}
\caption{Dependence of temperature of the trapped $^{39}\textup{K}$ atom cloud on the trap depth of the optical dipole trap. Solid red line represents a fit using the function $(1+Ax+Bx^2)^{1/2} - 1$, with A=4.45, B=0.0340, which becomes linear at the higher temperatures.} 
\label{Figure}
\end{figure}

We assume that the averaged  thermal energy difference $\langle \Delta U_{th}\rangle=\frac{3}{2}k_{B} T$ and perform thermal averaging of Eq. (4) to obtain an expression for the dynamic molecular polarizability in terms of the atomic polarizability and our measured gradient of the Feshbach position shift $\langle \Delta B\rangle/ U_{av}$:
\hspace*{0.8cm}

\begin{equation}
\alpha_{mol}/\alpha_{at}=2+\delta \mu\frac{\langle\Delta B\rangle}{U_{av}} -\frac{3}{2} \frac{k_B T}{U_{av}}.
\end{equation}

\noindent The cloud temperature is a nonlinear function of the ODT trap depth which was fitted to the experimental measurements of temperature (Fig. 10).

\section*{Acknowledgments}
We thank Andrea Simoni (University of Rennes) for providing Figs. 1 and 2 and for many fruitful discussions regarding interpretation of the results; Romain Vexiau (CNRS, Orsay) for providing the $\textup{K}_2$ $b{^3\Sigma_g^+}$ potential data; and Andrey Stolyarov (Moscow State University) for providing preliminary calculations of $\textup{K}_2$ molecular energy levels. We also thank Allessio Ciamei (LENS, Florence), Krzysztof Sacha (Jagiellonian University, Krakow), Hanns-Christoph N\"agerl (Innsbruck), Manuele Landini (Innsbruck) and Charly Beulenkamp (Innsbruck) for stimulating discussions on $^{39}\textup{K}$ Feshbach resonances; Arpana Singh, Chamali Gunawardana, and Satoshi Tojo (Chuo University, Tokyo) for contributions to the
early stages of the experiment; and Kanishk Gokul for
preparing Fig. 8. This research was funded
by the Australian Research Council Discovery Project
Grants DP190100815 and DP240101590.
\section*{Data Availability}
The data supporting the findings of the article are openly available \textcolor{red}{\cite{data}}.


\bibliography{references.bib}
\end{document}